\documentclass[aps,prd,twocolumn,nofootinbib,longbibliography]{revtex4-2}

\usepackage[utf8]{inputenc}
\usepackage[T1]{fontenc}
\usepackage{amsmath,amssymb,amsfonts,mathtools}
\usepackage{bm}
\usepackage{graphicx}
\usepackage{xcolor}
\usepackage{hyperref}
\usepackage{booktabs}
\usepackage{array}[=2016-10-06]
\usepackage{tabularx}
\hypersetup{colorlinks=true,linkcolor=blue,citecolor=blue,urlcolor=blue}

\begin{document}

\title{Energetic Ceilings and Benchmarks of Astrophysical Gravitational-Wave Backgrounds}

\author{Chiara M. F. Mingarelli}
\affiliation{Department of Physics, Yale University, New Haven, 06520, CT, USA}
\email{chiara.mingarelli@yale.edu}

\date{\today}

\begin{abstract}
\noindent Every astrophysical gravitational-wave background (GWB) is limited by the rest mass its sources can convert into gravitational radiation.  We combine the available source mass with the merger factor, radiative efficiency, and merger epochs for six GWB populations: supermassive black hole (SMBH) binaries (SMBHBs), extreme-mass-ratio inspirals (EMRIs), intermediate-mass black holes (IMBHs) inspiralling into SMBHs, stellar mass binary black holes (sBBHs), binary neutron stars (BNSs), and Population~III (Pop~III) binary black holes. For each, we calculate a GWB benchmark using a set of fiducial inputs and an upper limit. For SMBHBs, a recent local mass census from Liepold \& Ma (2024) together with a boost from repeated mergers calculated by Sato-Polito et al. (2024), allow us to compute a ceiling on the nanoHertz GWB. Anchoring the GWB to this census, and assuming one merger per SMBH, gives a benchmark of $A_{\rm bench} = 2.11^{+0.43}_{-0.32}\times10^{-15}$ at $f=1\,{\rm yr}^{-1}$. For the ceiling, we assume an infinite number of previous mergers and anchor the SMBHs to the same survey, yielding $A_{\rm ceil}=3.76^{+0.76}_{-0.56}\times10^{-15}$, consistent with all the published GWB amplitudes reported by pulsar timing arrays. Under this mass function and merger model, the measured GWB amplitude therefore does not require additional SMBH mass. In fact, when a GWB is measured, we can invert this logic to constrain the properties of its source population. For example, using the revised 15-year GWB amplitude from NANOGrav and the above SMBH census gives $f_{\rm merge}=1.11^{+1.35}_{-0.63}$, where $f_{\rm merge}=1$ means one merger per present-day remnant.  For the five other populations, comparable censuses do not yet exist, so we report upper limits using the best available source constraints, including current compact-binary merger rates. Finally, the six adopted upper limit cases yield a present-day gravitational-wave energy budget of $\rho_{\rm gw}^{(6)}=\rho_c\sum_i\int\Omega_i\,{\rm d}\ln f=3.33^{+1.32}_{-0.83}\times10^{-7}\rho_c$. A larger energy density would require more source mass, a larger merger factor, higher radiative efficiency, or later merger epochs. It could also point to additional source populations or new physics.
\end{abstract}

\maketitle

\section{Introduction}

The expanding network of gravitational-wave (GW) observatories now spans more than ten orders of magnitude in frequency. Pulsar timing arrays (PTAs) probe nanoHertz frequencies \cite{Agazie2023, Antoniadis2023, Reardon2023, Xu2023, Miles2025}, proposed missions such as $\mu$Ares target microhertz frequencies~\cite{Sesana2021}, and LISA will survey the millihertz sky~\cite{AmaroSeoane2017}.  The Big Bang Observer (BBO) targets the decihertz band~\cite{Crowder2005, Harry2006}, while ground-based detectors cover tens to thousands of hertz~\cite{LIGOScientific2015, Acernese2015, Akutsu2021, Punturo2010, Reitze2019}. Each frequency band can contain overlapping stochastic gravitational-wave backgrounds (GWBs) from multiple astrophysical populations. Attributing a measured signal among candidate populations therefore requires knowing which backgrounds are likely, and how large each one can be.

The concept that astrophysical backgrounds are bounded by a global mass budget is well established. \citet{Phinney2001} provided the general formalism linking the characteristic strain to the radiated energy spectrum of a population. For supermassive black hole (SMBH) binaries (SMBHBs), work by Sesana and others linked the expected background to galaxy assembly and SMBH demographics~\citep{Sesana2008,Sesana2013}. At the high-amplitude end, \citet{McWilliams14} maximized a forward model in which mergers alone drive late-time SMBH evolution. By contrast, \citet{ZhuCuiThrane2019} constructed a census-based GWB maximum without modeling that evolution: every present-day SMBH is treated as the remnant of one equal-mass merger.  \citet{satopolito2024} (hereafter SZQ24) extended this accounting to repeated mergers, showing that even an infinite merger hierarchy enhances \(h_c^2(f)\) at a particular frequency by only a factor of \(2.70\). In a complementary study of cosmological GWBs, \citet{2024arXiv240915572W} derived an integrated constraint for backgrounds already present at recombination.

These earlier results leave two problems open: how to compare astrophysical GWBs across the GW spectrum, and how to test whether a candidate population can supply a measured signal. We address both with a unified GWB framework which we apply to six populations. Building on Phinney's theorem~\cite{Phinney2001}, the framework tracks the available source mass, merger factor, radiative efficiency, and merger epochs. In reverse, it converts a measured background into a source-mass requirement that can be tested against an independent astrophysical census.

For SMBHBs, SZQ24 showed why this inverse question can be important. Using SMBH abundances inferred from local scaling relations, they found that even an upper limit on the GWB obtained by maximizing the merger history remained below the published PTA GWB measurements. Therefore, preserving the usual local SMBH mass density instead required a previously undercounted population of SMBHBs with total masses \(\gtrsim 3\times10^{10}\,M_\odot\). This is the ``missing SMBH'' question posed in their title: where are the supermassive black holes measured by PTAs? Whether this tension persists depends on the inferred abundance of SMBHs at the high-mass end of the local mass function.

The local SMBH census of \citet{Liepold2024} (hereafter LM24; also \cite{LiepoldMa2026Erratum}) allows us to calculate such a ceiling for the SMBHB GWB, and comment on this local high-mass tail. Inputs to the five upper limits include recent merger rates from LIGO--Virgo--KAGRA (LVK)~\cite{GWTC5pop}, the LM24 SMBH mass function, and the cosmic stellar mass density of \citet{MadauDickinson2014}.

The remainder of this paper is organized as follows. In Sec.~\ref{sec:energylimits}, we derive the energy conservation framework and its general scaling relations. Section~\ref{limitsOnAstroGWBs} applies the framework to SMBHBs, intermediate-mass-ratio inspirals (IMRIs) in active galactic nucleus (AGN) disks, in which intermediate-mass black holes (IMBHs) inspiral into SMBHs (AGN--IMRIs; see \cite{Mingarelli2026b}), extreme-mass-ratio inspirals (EMRIs), binary neutron stars (BNSs), binary black holes (BBHs) from Population~III (Pop~III) stars, and stellar mass BBHs (sBBHs). In Sec.~\ref{sec:astroCeilingOmegaGW}, we compare the benchmark energies and theoretical upper limits. We discuss our results and offer future outlooks in Sec.~\ref{sec:discussion}, conclude in Sec.~\ref{sec:conclusion}. Two appendices help the reader reconcile concepts and notation from this paper with others. Appendix~\ref{app:szq24-check} compares our SMBHB calculations with SZQ24, and Appendix~\ref{app:bbh-lvk-comparison} compares the sBBH calculation with the LVK GWB population models. Throughout this work we use units with $G=c=1$.

\section{A General Energetic Bound on Astrophysical GWBs}
\label{sec:energylimits}

Astrophysical GWBs are powered by sources that convert part of their rest-mass energy into gravitational radiation.  We describe the present-day background by
\begin{equation}
\begin{aligned}
\Omega_{\rm gw}(f)
&\equiv \frac{1}{\rho_c}
\frac{{\rm d}\rho_{\rm gw}}{{\rm d}\ln f} =\frac{2\pi^2}{3H_0^2}\,f^2h_c^2(f),
\end{aligned}
\label{eq:omega-strain}
\end{equation}
where ${\rm d}\rho_{\rm gw}/{\rm d}\ln f$ is the present-day GW energy density per logarithmic frequency and $\rho_c=3H_0^2/(8\pi )$ is the critical density ~\cite{Phinney2001,Mingarelli2019}.

Let $N(z){\rm d}z$ be the number of events per unit comoving volume between $z$ and $z+{\rm d}z$, and let the angle brackets denote an average over those events.  At observed frequency $f$, the frequency in the source's cosmic rest frame is $f_r=(1+z)f$.  At each frequency, Phinney’s theorem equates the present-day GW energy density to the redshifted energy emitted by all events:
\begin{equation}
\frac{{\rm d}\rho_{\rm gw}}{{\rm d}\ln f}
= \int_0^{z_{\rm max}}
\frac{N(z)}{1+z}
\left\langle f_r\frac{{\rm d}E}{{\rm d}f_r}\right\rangle_z
{\rm d}z.
\label{eq:cosmo-integral}
\end{equation}
When a population is specified by a comoving event-rate density $R(z)$ per unit source time, $N(z)=R(z)/[(1+z)H(z)]$, where  $H(z)=H_0\sqrt{\Omega_m(1+z)^3+\Omega_\Lambda}$, $H_0=67.4\ {\rm km\,s^{-1}\,Mpc^{-1}}$, $\Omega_m=0.3$ and $\Omega_\Lambda=0.7$~\cite{Planck2020}.

\subsection{A global mass-energy constraint}
\label{sec:globalMassEnergyConstraint}

Let $\rho_{\rm res}$ be the mass density in the source reservoir.  We define the source mass density $\rho_{\rm src}\equiv f_{\rm merge}\rho_{\rm res}$ as the total mass density entering GW-emitting events over the source history.  For comparable-mass mergers, this is the total binary mass; for EMRIs and IMRIs, it is the secondary mass.  The merger factor, $f_{\rm merge}\geq 0$, can exceed unity when the same mass participates in more than one merger.

Let $\langle E_{\rm gw}\rangle_z$ be the mean source-frame GW energy radiated per event at redshift $z$, and let $\epsilon_{\rm gw}$ be the fraction of $f_{\rm merge}\rho_{\rm res}$ radiated as GWs:
\begin{equation}
\int_0^{z_{\rm max}} N(z)\langle E_{\rm gw}\rangle_z\,{\rm d}z
=\epsilon_{\rm gw}f_{\rm merge}\rho_{\rm res}.
\end{equation}
Integrating Eq.~\eqref{eq:cosmo-integral} over observed frequency then gives
\begin{equation}
\begin{aligned}
\rho_{\rm gw}^{\rm tot}
&\equiv \rho_c\int_0^\infty
\Omega_{\rm gw}(f)\,{\rm d}\ln f \\
&=\int_0^{z_{\rm max}}\frac{N(z)}{1+z}
\langle E_{\rm gw}\rangle_z\,{\rm d}z \leq \epsilon_{\rm gw}f_{\rm merge}\rho_{\rm res} \, .
\end{aligned}
\label{eq:global-energy}
\end{equation}
Equation~\eqref{eq:global-energy} applies to both eccentric and circular binary systems, since the integrated bound depends on the total emitted energy. Intuitively, the factor $1/(1+z)\leq1$ shows that the present-day GW energy is maximized when all emission occurs at $z=0$.

One can rearrange the equation and interpret it as a way of setting a lower limit on the merger factor for a measured GWB:
\begin{equation}
f_{\rm merge}\geq
\frac{\rho_{\rm gw}^{\rm tot}}
{\epsilon_{\rm gw}\rho_{\rm res}}.
\end{equation}

For a specified population, we define $\mathcal H(f)$ such that it gives the fraction of the GW energy budget observed per logarithmic frequency interval:
\begin{equation}
\mathcal H(f)\equiv
\frac{1}{\epsilon_{\rm gw}f_{\rm merge}\rho_{\rm res}}
\frac{{\rm d}\rho_{\rm gw}}{{\rm d}\ln f} \, ,
\label{eq:history-spectrum-factor}
\end{equation}
and Eq.~\eqref{eq:global-energy} implies $\int_0^\infty\mathcal H(f)\,{\rm d}\ln f\leq1$. Combining Eqs.~\eqref{eq:omega-strain} and \eqref{eq:history-spectrum-factor} gives
\begin{equation}
\resizebox{0.98\columnwidth}{!}{$\displaystyle
h_c(f_{\rm ref})=
\left(\frac{3H_0^2}{2\pi^2f_{\rm ref}^2\rho_c}\right)^{1/2}
\rho_{\rm res}^{1/2}f_{\rm merge}^{1/2}
\epsilon_{\rm gw}^{1/2}\mathcal H(f_{\rm ref})^{1/2}.
$}
\label{eq:universal-scaling}
\end{equation}
We use Eq.~\eqref{eq:universal-scaling} for all six populations, and three labels for the results.  A \emph{benchmark} is the result for a set of inputs into Eq.~\eqref{eq:universal-scaling}.  An \emph{upper limit} is the largest value within a particular model's ranges. Unlike an upper limit, a \emph{ceiling} is anchored to an independent measurement of the population’s mass reservoir. It is the maximum strain at a given frequency consistent with that census or mass reservoir measurement. Currently only the GWB from SMBHBs has a ceiling; the other five results are upper limits.

\subsection{Comparing GW energy across frequency bands}
\label{sec:integrated-energies}

To compare source populations across frequency bands, we use the total GW energy density of each population $i$:
\begin{equation}
\begin{aligned}
\mathcal E_i
&\equiv  \!\! \int_0^\infty \Omega_i(f)\,{\rm d}\ln f
=\frac{\rho_{{\rm gw},i}^{\rm tot}}{\rho_c} \\
&= \frac{\epsilon_{{\rm gw},i}f_{{\rm merge},i}
\rho_{{\rm res},i}}{\rho_c} \!\!
\int_0^ \infty \!\!\! \mathcal H_i(f)\,{\rm d}\ln f \leq \!
\frac{\epsilon_{{\rm gw},i}f_{{\rm merge},i}
\rho_{{\rm res},i}}{\rho_c} \,
\end{aligned}
\label{eq:channel-integrated-energy}
\end{equation}
which follows from Eq.~\eqref{eq:history-spectrum-factor} and Eq.~\eqref{eq:global-energy}. We can evaluate Eq.~\eqref{eq:channel-integrated-energy} either by integrating $\Omega_i(f)$ or by summing the energy of the events.

The innermost stable circular orbit (ISCO) of a Schwarzschild black hole is
\begin{equation}
f_{\rm ISCO}=\frac{1}{6^{3/2}\pi M} \, .
\label{eq:f_isco_schwarzschild}
\end{equation}
Integrating the leading order circular inspiral spectrum to the ISCO gives the energy radiated by one binary, $E_{\rm insp}=\eta M/12$, where $M=M_1+M_2$ and $\eta=q/(1+q)^2$~\citep{1963PhRv..131..435P,1964PhRv..136.1224P,Bardeen1972}. Substituting this into Eq.~\eqref{eq:global-energy} gives
\begin{equation}
\begin{aligned}
\mathcal E_{\rm insp}
=\frac{f_{\rm merge}\rho_{\rm res}}{\rho_c}
\left\langle\frac{\eta}{12(1+z)}\right\rangle_{\rm M} \, ,
\end{aligned}
\label{eq:circular-inspiral-energy}
\end{equation}
where $\langle X\rangle_M\equiv\sum_k M_kX_k/\sum_kM_k$, so each merger is weighted by its total mass $M$. The integrated energies \(\mathcal E_i\) put GWBs in different frequency bands on a common energy scale, allowing us to compare the total GW energy in each one. In Sec.~\ref{sec:astroCeilingOmegaGW}, we compare the GW energies from each GWB to see which populations contribute most.

\section{Applications to GWBs}
\label{limitsOnAstroGWBs}

We use Eq.~\eqref{eq:universal-scaling} to calculate benchmarks and upper limits for six GWBs.  We begin with SMBHBs (Sec.~\ref{sec:SMBHB-PTA-Band}) and AGN--IMRIs (Sec.~\ref{sec:IMBH-SMBHcapture}), followed by EMRIs (Sec.~\ref{sec:EMRIs}), BNSs (Sec.~\ref{sec:BNS}), Pop~III BBHs (Sec.~\ref{sec:POPIII}), and sBBHs (Sec.~\ref{sec:stellarBBH}).  The double-white-dwarf (DWD) signals in Sec.~\ref{sec:DWD} are shown only for context. For each population, we estimate the available mass, how often that mass takes part in relevant mergers, what fraction becomes GW energy, and when and at what frequencies that energy is emitted, which define the benchmarks.

The three rate-based calculations use version 5.0 of the Gravitational-Wave Transient Catalog (GWTC-5.0) merger rates in Table~\ref{tab:lvk-rates}~\cite{GWTC5pop}. Their 90\% rate intervals are propagated to both $h_c^{\rm bench}$ and $\mathcal E_i$ in Table~\ref{tab:population_params}, with all other inputs fixed. The equations below give scaling relations evaluated at the median rates; the upper limit equations use the fixed rates in the upper limit column of Table~\ref{tab:lvk-rates}. We denote upper limit strains by $h_c^{\rm UL}$. Only the SMBHB calculation has an independent census of the relevant remnant population, so only the GWB from SMBHBs has a ceiling. Figure~\ref{fig:gwb-benchmarks} and Table~\ref{tab:population_params} show these benchmarks, Table~\ref{tab:upper_limits} lists the SMBHB ceiling and five upper limits, and Sec.~\ref{sec:astroCeilingOmegaGW} compares integrated GW energies.

\begin{table}[t!]
\centering
\caption{\textbf{GWTC-5.0 merger-rate inputs.} Rates are in $\mathrm{Gpc}^{-3}\,\mathrm{yr}^{-1}$~\cite{GWTC5pop}.  Benchmarks rates give the median and 90\% credible interval from the Default-BBH and FullPop models for BBHs and BNSs, respectively.  The upper limit rates are the larger endpoints across FullPop and PixelPop: PixelPop for BBHs and FullPop for BNSs.}
\label{tab:lvk-rates}
\small
\setlength{\tabcolsep}{3pt}
\renewcommand{\arraystretch}{1.05}
\begin{tabular}{@{}lcc@{}}
\toprule
\textbf{Population} & \textbf{Benchmark rate} & \textbf{Upper limit rate} \\
\midrule
BBH ($z=0.2$) & $30.4^{+6.5}_{-5.4}$ & $49.4$ \\
BNS ($z=0$) & $59.3^{+95.4}_{-43.9}$ & $154.7$ \\
\bottomrule
\end{tabular}
\end{table}

\subsection{Constraints on Supermassive Populations}
\label{sec:SMBHconstraints}

\subsubsection{SMBHBs (PTA Band)}
\label{sec:SMBHB-PTA-Band}

\label{sec:smbh_density}
LM24 derive the local SMBH mass function by convolving the $z=0$ galaxy stellar mass function, calibrated with the MASSIVE survey~\citep{Ma2014}, with the $M_\bullet$--$M_*$ relation~\citep{MM13,Gu2022,Liepold2024}.  Their mass function~\citep{LiepoldMa2026Erratum} gives
\begin{equation}
\rho_\bullet
=\left(1.8^{+0.8}_{-0.5}\right)\times10^6\,
M_\odot\,{\rm Mpc}^{-3},
\label{eq:rho_bullet_result}
\end{equation}
where the quoted range is a 90\% confidence interval.  This density is higher than earlier estimates based on galaxy velocity dispersions or quasar luminosity functions and the Soltan argument~\citep{Yu2002,Shankar2009,Soltan1982,Marconi2004,Shen2020}. The difference is driven mainly by the larger number of galaxies at the high-mass end of the LM24 stellar mass function.

At fixed total SMBH mass density, a population with more massive binaries produces a stronger GWB because $h_c^2\propto\mathcal{M}_c^{5/3}$, where $\mathcal{M}_c=(M_1M_2)^{3/5}(M_1+M_2)^{-1/5}$ is the chirp mass~\cite{JaffeBacker2003}.  Thus, Eq.~\eqref{eq:rho_bullet_result} gives the total available SMBH mass, while the LM24 mass function gives the abundance of SMBHs in each mass range. For comparison, SZQ24 use a velocity dispersion function (VDF)-based SMBH mass function and assume one generation of mergers in which every present-day SMBH is a remnant. Their one-generation construction is the $f_{\rm merge}=1$ benchmark in our notation.  Appendix~\ref{app:szq24-check} translates and compares the two calculations in detail.

We use the LM24 census to calculate the benchmark and physical ceiling. For our benchmark calculation, we use the LM24 mass function and their distributions $p(q)\propto q^2$ over $0.1\leq q\leq1$ and $p(z)\propto z\exp[-(z/0.5)^2]$ for $z\geq0$, setting $f_{\rm merge}=1$. Using the same remnant bookkeeping, we identify each present-day remnant mass with the total mass of its progenitor binary; this neglects GW mass loss and later accretion. For this channel, $\rho_{\rm res}=\rho_\bullet$. We truncate each inspiral at the ISCO of a Schwarzschild black hole~\cite{Bardeen1972}, Eq. \eqref{eq:f_isco_schwarzschild}.

We use $\epsilon_{\rm gw}=0.02$, while the LM24 population model determines $\mathcal H_{\rm LM24}$ at $1\,{\rm yr}^{-1}$. Using the mass density from Eq.~\eqref{eq:rho_bullet_result}, Eq.~\eqref{eq:universal-scaling} becomes:
\begin{widetext}
\begin{equation}
h_c^{\mathrm{bench}}
=2.11\times10^{-15}\left(\frac{\rho_\bullet}
{1.80\times10^6\,M_\odot\,{\rm Mpc}^{-3}}\right)^{\!1/2}
\left(\frac{f_{\rm merge}}{1}\right)^{\!1/2}\!\!
\left(\frac{\epsilon_{\rm gw}}{0.02}\right)^{\!1/2}\!\!
\left(\frac{\mathcal H}{\mathcal H_{\rm LM24}}\right)^{\!1/2} \, ,
\label{eq:smbhb-scaling}
\end{equation}
\end{widetext}
consistent with the published LM24 value of $2.0\times10^{-15}$.

\begin{figure}[t]
\centering
\includegraphics[width=0.9\columnwidth]{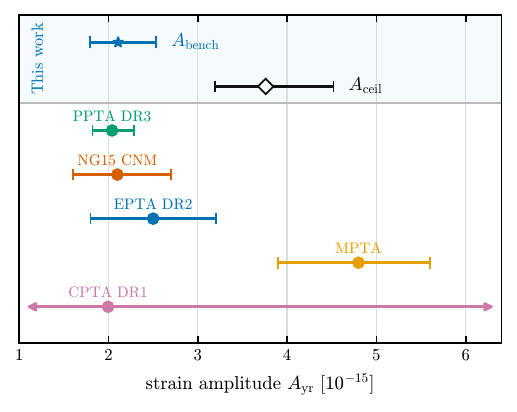}
\vspace{-.5\baselineskip}
\caption{\textbf{PTA amplitude comparison.} Evaluating Eq.~\eqref{eq:universal-scaling} for SMBHBs gives our benchmark, $h_c^{\rm bench}=2.11^{+0.43}_{-0.32}\times10^{-15}$ and ceiling $h_c^{\rm ceil}=3.76^{+0.76}_{-0.56}\times10^{-15}$.  The benchmark overlaps measurements from the Parkes PTA third data release (PPTA DR3), the NANOGrav 15-year custom noise model (NG15 CNM), and the European PTA second data release (EPTA DR2), the MeerKAT PTA (MPTA), and the Chinese PTA first data release (CPTA DR1). With the exception of the latter, they ~\citep{Reardon2023,Larsen2026,Agarwal2026,Antoniadis2023,Miles2025,Xu2023}, fit comfortably within the ceiling.}
\label{fig:pta-amplitudes}
\end{figure}

To compute the physical SMBHB ceiling, we keep the LM24 mass function fixed, set $q=1$ and $z\to0$, and include the complete SZQ24 merger hierarchy. These factors respectively increase the benchmark $h_c^2$ by $1.05\times1.12\times2.70=3.18$.  The physical ceiling is therefore
\begin{equation}
\begin{aligned}
h_c^{\rm ceil}(1\,{\rm yr}^{-1})
&=\sqrt{3.18}\,h_c^{\rm bench}(1\,{\rm yr}^{-1}) =3.76\times10^{-15}.
\end{aligned}
\label{eq:smbhb-repeated-ceiling}
\end{equation}
The leading order inspiral strain used for the ceiling is insensitive to spins, however, spins do matter for the late inspiral, merger, and ringdown (IMR) for other binary populations.

Importantly, the SMBH census and a GWB measurement can be used to constrain $f_{\rm merge}$. In this approach, the LM24 amplitude $A_{\rm LM24}$ corresponds to $f_{\rm merge}=1$. Let $A_{\rm CNM}$ denote the revised NANOGrav 15-year custom noise model amplitude. For the LM24 mass ratio and redshift distributions, Eq.~\eqref{eq:smbhb-scaling} gives
\begin{equation}
f_{\rm merge}
=\left(\frac{A_{\rm CNM}}{A_{\rm LM24}}\right)^2.
\label{eq:fmerge-propagation}
\end{equation}

\begin{figure}[t!]
\centering
\includegraphics[width=\columnwidth]{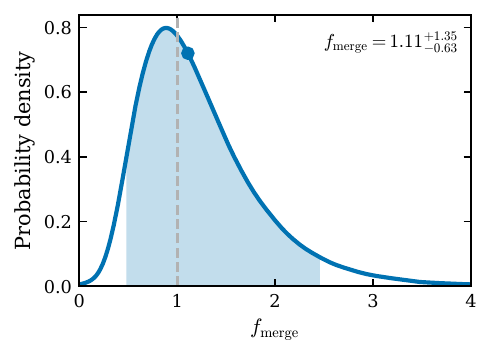}
\vspace{-1.5\baselineskip}
\caption{\textbf{SMBHB merger factor from the GWB.}  Posterior for $f_{\rm merge}$ from Eq.~\eqref{eq:fmerge-propagation}, using the NANOGrav amplitude and the LM24 Appendix B reconstruction. The blue dot marks the median, the shaded region is the 90\% credible interval, and the light-gray dashed line marks $f_{\rm merge}=1$, corresponding to one merger per present-day SMBH remnant.}
\label{fig:fmerge-posterior}
\end{figure}

Estimating its uncertainty, however, requires their amplitude distributions. We approximate the fixed-$\gamma=13/3$ NANOGrav distribution with a two-sided Gaussian in $\ln A_{\rm CNM}$ matched to its published median and 90\% interval~\citep{Larsen2026,Agarwal2026}, and reconstruct the LM24 distribution from their Appendix~B ~\citep{Liepold2024,LiepoldMa2026Erratum}. Applying Eq.~\eqref{eq:fmerge-propagation} to independent draws from these distributions, with a prior uniform in $\log_{10}f_{\rm merge}$ over $-3\leq \log_{10}f_{\rm merge}\leq1$, gives
\begin{equation}
f_{\rm merge}
=1.11^{+1.35}_{-0.63}\qquad(90\%).
\label{eq:nanograv-fmerge}
\end{equation}
The 90\% interval includes $f_{\rm merge}=1$, making the GWB amplitude consistent with one merger per present-day remnant.

\begin{table*}[ht!]
\centering
\caption{\textbf{Benchmarks for the six modeled astrophysical GWBs.} We list the available mass, source fraction, GW efficiency, characteristic strain at the stated reference frequency, and the dimensionless GW energy density defined in Eq.~\eqref{eq:channel-integrated-energy}.  The model column summarizes the source history and spectrum. IMR denotes inspiral--merger--ringdown.  The AGN--IMRI, BNS, and sBBH benchmarks use a 10 Gyr constant rate history.  Their rate inputs are listed in Table~\ref{tab:lvk-rates}.  Mass densities are in $M_\odot\,{\rm Mpc}^{-3}$.  The SMBHB energy is obtained from the global mass accounting in Eq.~\eqref{eq:circular-inspiral-energy}; the other energies are integrated over their modeled spectra using Eq.~\eqref{eq:channel-integrated-energy}.}
\label{tab:population_params}
\small
\setlength{\tabcolsep}{2pt}
\renewcommand{\arraystretch}{0.90}

\begin{tabularx}{\textwidth}{@{}>{\raggedright\arraybackslash}p{0.10\textwidth} >{\raggedright\arraybackslash}p{0.14\textwidth} >{\raggedright\arraybackslash}p{0.16\textwidth} @{\hspace{6pt}}>{\centering\arraybackslash}p{0.065\textwidth}@{\hspace{6pt}} >{\raggedright\arraybackslash}p{0.15\textwidth} @{\hspace{2pt}}>{\raggedright\arraybackslash}p{0.13\textwidth}@{\hspace{2pt}} >{\raggedright\arraybackslash}X@{}}
\toprule
\textbf{Channel} & \textbf{Available mass} & \textbf{Source fraction} & $\bm{\epsilon_{\rm gw}}$ & $\bm{h_c^{\mathrm{bench}}(f_{\rm ref})}$ & $\bm{\mathcal E_i}$ & \textbf{Model} \\
\midrule
SMBHB & $\rho_\bullet=1.80\times10^6$  & $f_{\rm merge}=1$ & $\simeq0.02$ & \begin{tabular}[t]{@{}l@{}}$2.11^{+0.43}_{-0.32}\times10^{-15}$\\$f_{\rm ref}=1\,{\rm yr}^{-1}$\end{tabular} & $2.02^{+0.90}_{-0.56}\times10^{-7}$ &  LM24 census; inspiral to ISCO (Sec.~\ref{sec:SMBHB-PTA-Band}) \\
AGN--IMRI & $\rho_{\rm BBH}=1.82\times10^4$ & $f_{\rm AGN}\kappa_{\rm IM}=0.03$ & $0.05$ & \begin{tabular}[t]{@{}l@{}}$1.35^{+0.14}_{-0.13}\times10^{-21}$\\$f_{\rm ref}=3\,{\rm mHz}$\end{tabular} & $1.48^{+0.32}_{-0.26}\times10^{-10}$ & \begin{tabular}[t]{@{}l@{}}IMBH grows in AGN disk\\(Sec.~\ref{sec:IMBH-SMBHcapture})\end{tabular} \\
EMRI & $\rho_{\rm nuc}=6.30\times10^5$ & $f_{\rm merge}=10^{-4}$ & $0.05$ & \begin{tabular}[t]{@{}l@{}}$3.48\times10^{-22}$\\$f_{\rm ref}=0.01\,{\rm Hz}$\end{tabular} & $2.50\times10^{-11}$ & Babak et al. M1; $z=0$ inspiral (Sec.~\ref{sec:EMRIs}) \\
BNS & $\rho_\star=6.30\times10^8$ & $f_{\rm merge}=2.63\times10^{-6}$ & $0.02$ & \begin{tabular}[t]{@{}l@{}}$4.28^{+2.63}_{-2.10}\times10^{-24}$\\$f_{\rm ref}=0.1\,{\rm Hz}$\end{tabular} & $1.80^{+2.90}_{-1.33}\times10^{-10}$ & Inspiral to contact (Sec.~\ref{sec:BNS}) \\
Pop~III & $\rho_{\rm III}=7.33\times10^4$ & $f_{\rm merge}=0.01$ & $0.05$ & \begin{tabular}[t]{@{}l@{}}$6.80\times10^{-24}$\\$f_{\rm ref}=0.1\,{\rm Hz}$\end{tabular} & $3.33\times10^{-11}$ & \begin{tabular}[t]{@{}l@{}}Liu--Bromm, no delay; IMR\\(Sec.~\ref{sec:POPIII})\end{tabular} \\
sBBH & $\rho_\star=6.30\times10^8$ & $f_{\rm merge}=1.05\times10^{-5}$ & $0.05$ & \begin{tabular}[t]{@{}l@{}}$4.03^{+0.41}_{-0.38}\times10^{-25}$\\$f_{\rm ref}=25\,{\rm Hz}$\end{tabular} & $1.79^{+0.38}_{-0.32}\times10^{-9}$ & Nonspinning IMR (Sec.~\ref{sec:stellarBBH}) \\
\bottomrule
\end{tabularx}
\end{table*}

\subsubsection{The AGN--IMRI GWB (LISA Band)}
\label{sec:IMBH-SMBHcapture}

AGN disks have long been studied as environments where stellar mass black holes can migrate, accrete, and merge repeatedly, with some remnants growing into IMBHs~\cite{McKernan2012,mckernan2014,Bellovary2016,Tagawa2020,Grishin2024}; see \citet{FordMcKernan2025} for a recent review. Some of these IMBHs later inspiral into the central SMBH, and the incoherent superposition of this cosmic SMBH-IMBH  population produces the AGN--IMRI GWB~\cite{Mingarelli2026b}. For this channel, $\rho_{\rm src}$ counts the IMBH mass entering the final IMBH--SMBH inspirals, not the SMBH mass:
\begin{equation}
f_{\rm merge}\rho_{\rm res}
=\rho_{\rm src}^{\rm AGN\text{-}IMRI}
=\rho_{\rm IMR} \, ,
\label{eq:fimr-def}
\end{equation}
where $\rho_{\rm IMR}$ is the total mass density in the IMBHs.  We define the processed sBBH mass density as $\rho_{\rm BBH,proc}\equiv R_{\rm BBH}\langle M_{\rm BBH}\rangle T_{\rm eff}$, where $R_{\rm BBH}$ is the total comoving sBBH merger rate density listed in Table~\ref{tab:lvk-rates}, $\langle M_{\rm BBH}\rangle$ is the representative total binary mass per merger, and $T_{\rm eff}$ is the lookback interval over which the source-frame rate is held constant. The companion paper calculates $\rho_{\rm IMR}$ from these quantities:
\begin{widetext}
\begin{equation}
\rho_{\rm IMR}
=f_{\rm AGN}\kappa_{\rm IM}\rho_{\rm BBH,proc}
=547.2\,M_\odot\,{\rm Mpc}^{-3}
\left(\frac{\kappa_{\rm IM}}{0.30}\right)\!\!
\left(\frac{f_{\rm AGN}}{0.10}\right)\!\!
\left(\!\frac{R_{\rm BBH}}{30.4\,{\rm Gpc}^{-3}{\rm yr}^{-1}}\!\right)\!\!
\left(\frac{\langle M_{\rm BBH}\rangle}{60\,M_\odot}\right)\!\!
\left(\frac{T_{\rm eff}}{10\,{\rm Gyr}}\right)\!.
\label{eq:agn-imri-source-density}
\end{equation}
\end{widetext}

The factor $f_{\rm AGN}$ selects the sBBH merger mass processed in AGN disks, while $\kappa_{\rm IM}$ selects the fraction of that mass retained in IMBHs that later inspiral into central SMBHs.  Comparing Eqs.~\eqref{eq:fimr-def} and \eqref{eq:agn-imri-source-density}, we can write $\rho_{\rm res}=\rho_{\rm BBH,proc}$ and $f_{\rm merge}=f_{\rm AGN}\kappa_{\rm IM}$ for this channel.

The benchmark value of $\rho_{\rm IMR}$ follows from Eq.~\eqref{eq:agn-imri-source-density}.  Its uncertainty comes from the 90\% credible interval on $R_{\rm BBH}$ in Table~\ref{tab:lvk-rates}; the other scaling values are held fixed. The choice $f_{\rm AGN}=0.1$ assigns 10\% of sBBH mergers to AGN disks.  The corresponding rate for this channel is $(3.04^{+0.65}_{-0.54})\,{\rm Gpc}^{-3}\,{\rm yr}^{-1}$, which is within the range predicted by \citet{Tagawa2020}.

We set the high-frequency cutoff using the smallest central SMBH in the model, $M_\bullet=10^5\,M_\odot$~\cite{Greene2020}. For a Schwarzschild black hole, this gives $f_{\rm max}=0.04\,{\rm Hz}$. For the 10 Gyr rate history summarized in Table~\ref{tab:population_params}, let $\mathcal H_{10}$ denote $\mathcal H(3\,{\rm mHz})$ from Eq.~\eqref{eq:history-spectrum-factor}. The benchmark is therefore:
\begin{equation}
\begin{aligned}
&h_c^{\mathrm{bench}}(3\,{\rm mHz})=1.35\times10^{-21}\\
&\times\!\left(\frac{\rho_{\rm IMR}}
{5.47\times10^2\,M_\odot\,{\rm Mpc}^{-3}}\right)^{\!1/2}\!\!
\left(\frac{\epsilon_{\rm gw}}{0.05}\right)^{\!1/2}\!\!
\left(\frac{\mathcal H}{\mathcal H_{10}}\right)^{\!1/2}\mathrlap{.}
\end{aligned}
\label{eq:agn-imri-benchmark}
\end{equation}

To calculate the AGN--IMRI upper limit, we use the sBBH rate in the upper limit column of Table~\ref{tab:lvk-rates} and take the largest stated value of each remaining input in Eq.~\eqref{eq:agn-imri-source-density}:
\begin{equation}
\begin{gathered}
\kappa_{\rm IM}=1,\qquad f_{\rm AGN}=1,\\
\langle M_{\rm BBH}\rangle=150\,M_\odot,\qquad
T_{\rm eff}=13\,{\rm Gyr}.
\end{gathered}
\label{eq:agn-imri-ul-inputs}
\end{equation}
The choices $f_{\rm AGN}=\kappa_{\rm IM}=1$ assign every sBBH merger to the AGN channel and assume that all of its mass is retained in IMBHs that later inspiral into central SMBHs.  These upper values give $\rho_{\rm IMR}=9.63\times10^4\,M_\odot\,{\rm Mpc}^{-3}$. The efficiency $\epsilon_{\rm gw}=0.05$ remains at its benchmark value. Let $\mathcal H_{13}$ denote $\mathcal H(3\,{\rm mHz})$ from Eq.~\eqref{eq:history-spectrum-factor} for the corresponding 13 Gyr population.  Equation~\eqref{eq:universal-scaling} then gives

\begin{equation}
\begin{aligned}
&h_c^{\mathrm{UL}}(3\,{\rm mHz})=1.73\times10^{-20}\\
&\times\!\left(\frac{\rho_{\rm IMR}}
{9.63\times10^4\,M_\odot\,{\rm Mpc}^{-3}}\right)^{\!1/2}\!\!
\left(\frac{\epsilon_{\rm gw}}{0.05}\right)^{\!1/2}\!\!
\left(\frac{\mathcal H}{\mathcal H_{13}}\right)^{\!1/2}\mathrlap{.}
\end{aligned}
\label{eq:agn-imri-ul}
\end{equation}

This upper limit combines that rate and four upper input values chosen independently; no single population is known to realize all five at once. Upcoming forward models can improve this estimate by determining which combinations of these inputs occur together.

\begin{figure*}
\centering
\includegraphics[width=\linewidth]{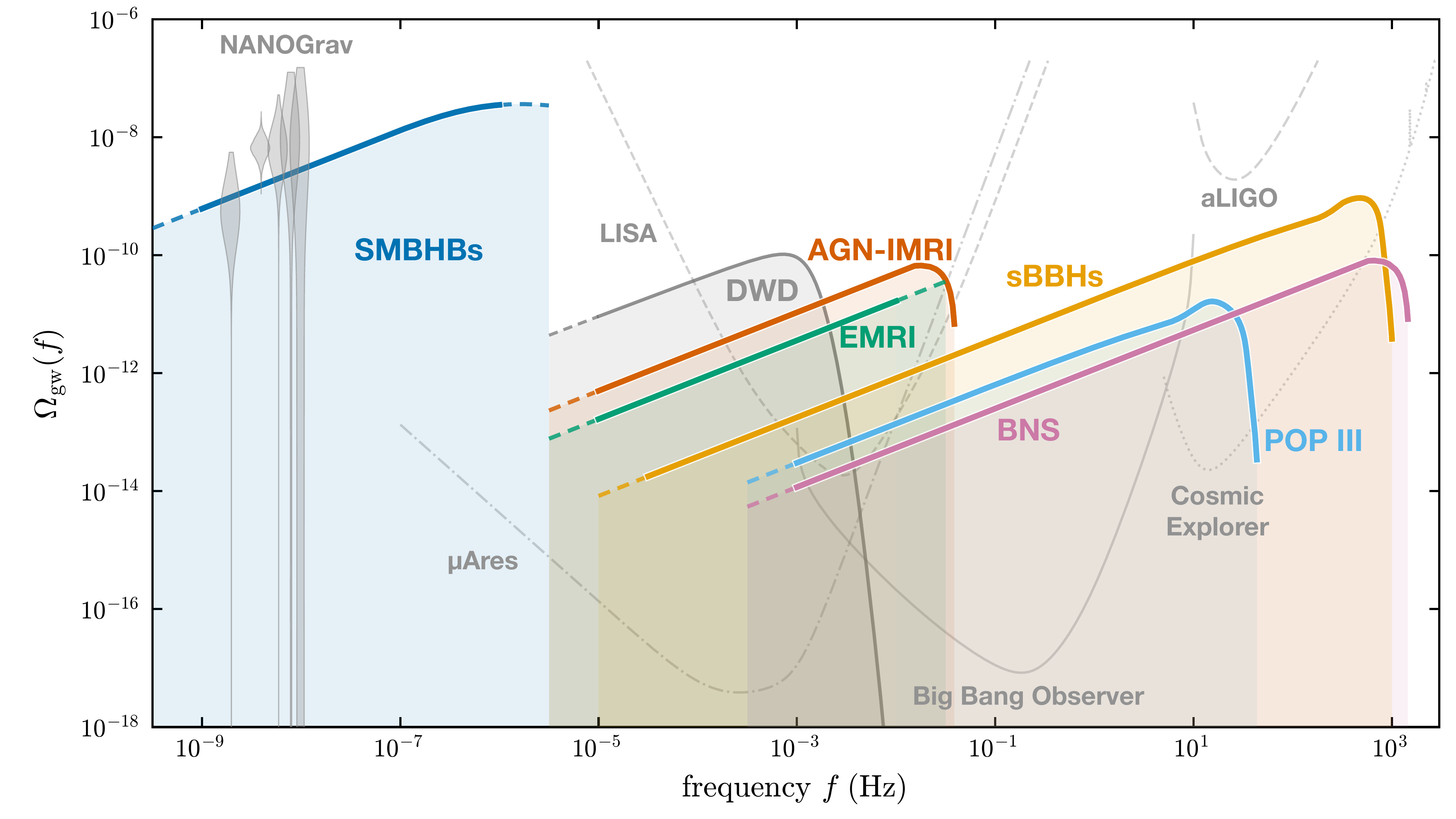}
\vspace{-1.5\baselineskip}
\caption{\textbf{Astrophysical GWB benchmarks across frequency.} Colored curves show the six central benchmark spectra defined in Table~\ref{tab:population_params} and Sec.~\ref{limitsOnAstroGWBs}. Colored shading is visual only and dashed segments are guide extensions of the populations. Curve height gives $\Omega_{\rm gw}$, the GW energy density per logarithmic frequency interval; the total energy is $\mathcal E_i=\int\Omega_i(f)\,{\rm d}\ln f$. The integrated values in Table~\ref{tab:population_params} show that the SMBHB benchmark contains about $10^2$ times more energy than the sBBH benchmark since much more mass is processed through SMBHB mergers. Light-gray violins show the revised NANOGrav 15-year free-spectrum posteriors in five frequency bins, converted to $\Omega_{\rm gw}$ ~\cite{Larsen2026,Agarwal2026}. The gray DWD curve is the unresolved Galactic foreground of \citet{Robson2019}. Gray detector curves assume 10 yr for $\mu$Ares, 4 yr for LISA, 5 yr for BBO, 2 yr of coincident Advanced-LIGO--Virgo operation at design sensitivity, and 1 yr for Cosmic Explorer ~\cite{Sesana2021,Robson2019,Schmitz2021, LIGOScientific2019Stochastic,Kuns2025CESensitivity}. These published sensitivity estimates use different conventions and are primarily shown to indicate band coverage. }
\label{fig:gwb-benchmarks}
\end{figure*}

\subsection{Constraints on Stellar Mass Populations}
\label{sec:stellarConstraints}
For EMRIs, BNSs, and sBBHs, we adopt $\rho_\star=5\times10^{-3}\rho_c =6.3\times10^8\,M_\odot\,{\rm Mpc}^{-3}$, the present-day cosmic stellar mass density inferred from the star-formation history of \citet{MadauDickinson2014}. Pop~III instead uses only the mass formed in metal-free stars.

Only a small fraction of $\rho_\star$ contributes to each background. We determine that fraction from the nuclear star cluster (NSC) abundance for EMRIs and from the measured merger rates for BNSs and sBBHs, see Table \ref{tab:lvk-rates}.

\subsubsection{EMRIs (LISA Band)}
\label{sec:EMRIs}
EMRIs form when stellar encounters in an NSC place a compact object on an orbital path leading to its capture by the central SMBH. Repeated gravitational encounters in the NSC gradually change the compact object's orbit, occasionally placing them on trajectories that lead to capture by the central SMBH~\cite{Gair2004,Hopman2005}.

Assuming every galaxy contains an NSC, integrating the median relation between NSC mass and host-galaxy mass~\cite{Neumayer2020} over the local galaxy stellar mass function~\cite{Baldry2012} gives $\rho_{\rm nuc}\simeq10^{-3}\rho_\star$. Scatter in the relation increases this estimate, while the fact that not every galaxy contains an NSC reduces it~\cite{SanchezJanssen2019,Neumayer2020}. Let $\rho_{\star,\rm bench}=6.30\times10^8\,M_\odot\,{\rm Mpc}^{-3}$ \citet{MadauDickinson2014}, we can therefore write
\begin{equation}
\begin{aligned}
\eta_{\rm NSC}&\equiv\frac{\rho_{\rm nuc}}{\rho_\star}=10^{-3},\\
\rho_{\rm nuc,bench}
&=\eta_{\rm NSC}\rho_{\star,\rm bench}
=6.30\times10^5\,M_\odot\,{\rm Mpc}^{-3}.
\end{aligned}
\label{eq:emri-reservoir}
\end{equation}
For EMRIs,
\begin{equation}
E_{\rm gw}^{\rm EMRI}
= \epsilon_{\rm gw}^{\rm EMRI}M_2\left[1+\mathcal{O}(q)\right],
\qquad q\equiv\frac{M_2}{M_1}\ll1,
\label{eq:emri-energy-scaling}
\end{equation}
Here $M_2$ is the captured object's mass, and $\epsilon_{\rm gw}^{\rm EMRI}$ depends on the primary spin and orbit~\cite{Hughes2001,Babak2017}. For this channel, $\rho_{\rm res}=\rho_{\rm nuc}$, $\rho_{\rm src}=\rho_{\rm src}^{\rm EMRI}$ is the mass in captured objects, and $f_{\rm merge}=f_{\rm merge}^{\rm EMRI} =\rho_{\rm src}^{\rm EMRI}/\rho_{\rm nuc}$ is the fraction of NSC mass captured in EMRIs.

The fiducial M1 model from \citet{Babak2017}, adopted by \citet{BonettiSesana2020}, predicts $1600$ EMRIs per year out to $z=4.5$, with $10\,M_\odot$ secondaries. Converting this all-sky rate to a mean comoving rate density out to $z=4.5$ and integrating it for $10$ Gyr gives $f_{\rm merge}^{\rm EMRI}\simeq10^{-4}$.  The resulting mass density in captured objects is
\begin{equation}
\begin{aligned}
&\rho_{\rm src}^{\rm EMRI}=f_{\rm merge}^{\rm EMRI}\rho_{\rm nuc} \\
&=63.0\,M_\odot\,{\rm Mpc}^{-3}\left(\frac{\rho_\star}{\rho_{\star,\rm bench}}\right)\!\!
\left(\frac{\eta_{\rm NSC}}{10^{-3}}\right)
\left(\frac{f_{\rm merge}^{\rm EMRI}}{10^{-4}}\!\!\right)\mathrlap{.}
\end{aligned}
\label{eq:emri-source-density}
\end{equation}
We use the total M1 rate to estimate the captured mass. Since its redshift distribution is unavailable, we evaluate the benchmark at $z=0$. At the adopted cutoff frequency, we define $\mathcal H_{\rm EMRI}\equiv\mathcal H(0.01\,{\rm Hz})$. With $\epsilon_{\rm gw}=0.05$, Eq.~\eqref{eq:universal-scaling} becomes
\begin{equation}
\begin{aligned}
&h_c^{\mathrm{bench}}(0.01\,{\rm Hz})=3.48\times10^{-22}\\
&\times\!\left(\frac{\rho_{\rm nuc}}{\rho_{\rm nuc,bench}}\right)^{\!1/2}
\left(\frac{f_{\rm merge}^{\rm EMRI}}{10^{-4}}\right)^{\!1/2}
\left(\frac{\epsilon_{\rm gw}}{0.05}\right)^{\!1/2}
\left(\frac{\mathcal H}{\mathcal H_{\rm EMRI}}\right)^{\!1/2}\mathrlap{.}
\end{aligned}
\label{eq:emri-benchmark}
\end{equation}
This benchmark reproduces the scale of the full population synthesis: the fiducial M1 background of \citet{BonettiSesana2020}, computed from the same rate model with all harmonics of the eccentric inspirals, reaches $h_c\simeq1.0\times10^{-21}$ at $3$\,mHz, comparable to the LISA sensitivity there, or $\simeq4.4\times10^{-22}$ at $0.01$\,Hz using $h_c\propto f^{-2/3}$, which is within about $30\%$ of Eq.~\eqref{eq:emri-benchmark}.

Predictions for EMRI rates vary by more than three orders of magnitude, and AGN disks may in fact increase them~\cite{Babak2017,BonettiSesana2020,Pan2021,Tagawa2020}. To set an upper limit, we use the stellar mass black holes available for capture in NSCs. Using a Salpeter distribution of stellar birth masses and $10\,M_\odot$ remnants, \citet{Babak2017} estimate that stellar mass black holes make up about $3\%$ of the stellar mass in galactic bulges. We apply the same fraction to NSCs and assume that every stellar mass black hole produces one EMRI, with no replenishment. This sets $f_{\rm merge}^{\rm EMRI}=0.03$. Holding the other benchmark inputs fixed, Eq.~\eqref{eq:universal-scaling} gives
\begin{equation}
\begin{aligned}
&h_c^{\mathrm{UL}}(0.01\,{\rm Hz})=6.02\times10^{-21}\\[-0.3ex]
&\times\left(\frac{\rho_{\rm nuc}}{\rho_{\rm nuc,bench}}\right)^{1/2}
\left(\frac{f_{\rm merge}^{\rm EMRI}}{0.03}\right)^{1/2}\left(\frac{\epsilon_{\rm gw}}{0.05}\right)^{1/2}
\left(\frac{\mathcal H}{\mathcal H_{\rm EMRI}}\right)^{1/2}.
\end{aligned}
\label{eq:emri-ul}
\end{equation}

\subsubsection{Binary Neutron Stars (decihertz Band)}
\label{sec:BNS}
For BNSs, we use the GWTC-5.0 local merger rate estimates listed in Table~\ref{tab:lvk-rates}. No new BNS candidates were reported through the second half of observing run 4 (O4b), leaving a broad allowed rate range~\cite{GWTC5,GWTC5pop}.

The upper frequency cutoff for BNS mergers is the contact (inspiral termination) frequency, which is governed by the neutron star equation of state~\cite{Read2013}. For a canonical $1.4+1.4\,M_\odot$ system, we use the cutoff frequency $f_{\rm max}=1.5$ kHz. For $\langle M_{\rm BNS}\rangle=2.8\,M_\odot$, a constant source-frame rate over $T_{\rm eff}=10$ Gyr reaches $z=1.66$. For BNSs, $\rho_{\rm res}=\rho_\star$, $\rho_{\rm src}^{\rm BNS}$ is the total binary mass entering mergers, and $f_{\rm merge}^{\rm BNS}=\rho_{\rm src}^{\rm BNS}/\rho_\star$ is this mass divided by the stellar mass density. The mass in BNS mergers is
\begin{equation}
\begin{aligned}
&\rho_{\rm src}^{\rm BNS}=f_{\rm merge}^{\rm BNS}\rho_\star
=1.66\times10^3\,M_\odot\,{\rm Mpc}^{-3}\\[-0.3ex]
&\times\!\left(\frac{R_{\rm BNS}}{59.3\,{\rm Gpc}^{-3}{\rm yr}^{-1}}\right)
\left(\frac{\langle M_{\rm BNS}\rangle}{2.8\,M_\odot}\right)
\left(\frac{T_{\rm eff}}{10\,{\rm Gyr}}\right)\mathrlap{.}
\end{aligned}
\label{eq:bns-source-density}
\end{equation}

Across the benchmark rate interval in Table~\ref{tab:lvk-rates}, Eq.~\eqref{eq:bns-source-density} gives $\rho_{\rm src}^{\rm BNS}=(1.66^{+2.67}_{-1.23})\times10^3\, M_\odot\,{\rm Mpc}^{-3}$ and $f_{\rm merge}^{\rm BNS}=(2.63^{+4.24}_{-1.95})\times10^{-6}$, with the mass and history fixed. As for SMBHBs, we include only the inspiral and set $\epsilon_{\rm gw}=0.02$~\cite{1963PhRv..131..435P,1964PhRv..136.1224P}. For this benchmark, let $\mathcal H_{\rm BNS}\equiv\mathcal H(0.1\,{\rm Hz})$, and Eq.~\eqref{eq:universal-scaling} gives:
\begin{equation}
\begin{aligned}
&h_c^{\mathrm{bench}}(0.1\,{\rm Hz})=4.28\times10^{-24}\\
&\times\left(\frac{\rho_\star}{\rho_{\star,\rm bench}}\right)^{1/2}
\left(\frac{f_{\rm merge}^{\rm BNS}}{2.63\times10^{-6}}\right)^{1/2}
\left(\frac{\epsilon_{\rm gw}}{0.02}\right)^{1/2}
\left(\frac{\mathcal H}{\mathcal H_{\rm BNS}}\right)^{1/2}.
\end{aligned}
\label{eq:bns-benchmark}
\end{equation}
For the upper limit, only the rate changes, see Table~\ref{tab:lvk-rates}, while $\rho_\star$ and $\mathcal H$ remain at their benchmark values.  Thus,
\begin{equation}
\begin{aligned}
&h_c^{\mathrm{UL}}(0.1\,{\rm Hz})=6.91\times10^{-24}\\
&\times\left(\frac{\rho_\star}{\rho_{\star,\rm bench}}\right)^{1/2}
\left(\frac{f_{\rm merge}^{\rm BNS}}{6.87\times10^{-6}}\right)^{1/2}
\left(\frac{\epsilon_{\rm gw}}{0.02}\right)^{1/2}
\left(\frac{\mathcal H}{\mathcal H_{\rm BNS}}\right)^{1/2}.
\end{aligned}
\label{eq:bns-ceiling}
\end{equation}

Since BNSs have small chirp masses, they evolve slowly through the low-frequency band: their residence time per logarithmic frequency interval scales as \(t_{\rm res}\equiv f/\dot f\propto \mathcal{M}_c^{-5/3}f^{-8/3}\). Consequently, many more low-mass signals overlap in the decihertz band than in the LVK band. Figure~\ref{fig:gwb-benchmarks} shows the total BNS signal before resolvable sources are removed; the remaining background depends on detector sensitivity and source subtraction~\cite{Rosado2011,Lawrence2023}.

\begin{table*}[ht!]
\centering
\caption{\textbf{Physical SMBHB strain ceiling and theoretical upper limits.} The $\mathcal E_i$ column gives the integrated energies summed in Eq.~\eqref{eq:integrated-reference-sum}.  For SMBHBs, this integral uses one $q=1$, $z\to0$ generation, is given by Eq.~\eqref{eq:circular-inspiral-energy}, and excludes the fixed-frequency factor $2.70$. The SMBHB intervals propagate the uncertainty in $\rho_\bullet$. The five upper limits are calculated from the upper limit inputs specified for each channel, while the rate-based cases use the upper limit rates in Table~\ref{tab:lvk-rates}.}
\label{tab:upper_limits}
\small
\setlength{\tabcolsep}{3pt}
\renewcommand{\arraystretch}{0.90}
\begin{tabularx}{\textwidth}{@{}>{\raggedright\arraybackslash}p{0.11\textwidth} >{\raggedright\arraybackslash}p{0.18\textwidth} >{\raggedright\arraybackslash}p{0.14\textwidth} >{\raggedright\arraybackslash}X@{}}
\toprule
\textbf{Channel} & $\bm{h_c^{\mathrm{ceil/UL}}(f_{\rm ref})}$ & $\bm{\mathcal E_i}$ & \textbf{Changes from the benchmark} \\
\midrule
SMBHB & \begin{tabular}[t]{@{}l@{}}$3.76^{+0.76}_{-0.56}\times10^{-15}$\\$f_{\rm ref}=1\,{\rm yr}^{-1}$\end{tabular} & \leavevmode$2.97^{+1.32}_{-0.83}\times10^{-7}$ & Set $q=1$ and $z\to0$; multiply fixed-frequency $h_c^2$ by $2.70$ for the mass-conserving hierarchy (Eq.~\eqref{eq:smbhb-repeated-ceiling}; Appendix~\ref{app:szq24-check}) \\
AGN--IMRI & \begin{tabular}[t]{@{}l@{}}$1.73\times10^{-20}$\\$f_{\rm ref}=3\,{\rm mHz}$\end{tabular} & $2.25\times10^{-8}$ & \leavevmode Use the sBBH rate in the upper limit column of Table~\ref{tab:lvk-rates} and the four upper inputs in Eq.~\eqref{eq:agn-imri-ul-inputs}; $\rho_{\rm IMR}=9.63\times10^4\,M_\odot\,{\rm Mpc}^{-3}$ (Eq.~\eqref{eq:agn-imri-ul}) \\
EMRI & \begin{tabular}[t]{@{}l@{}}$6.02\times10^{-21}$\\$f_{\rm ref}=0.01\,{\rm Hz}$\end{tabular} & $7.50\times10^{-9}$ & Set $f_{\rm merge}=0.03$ for an unreplenished NSC (Eq.~\eqref{eq:emri-ul}) \\
BNS & \begin{tabular}[t]{@{}l@{}}$6.91\times10^{-24}$\\$f_{\rm ref}=0.1\,{\rm Hz}$\end{tabular} & $4.70\times10^{-10}$ & \leavevmode Use the BNS rate in the upper limit column of Table~\ref{tab:lvk-rates} (Eq.~\eqref{eq:bns-ceiling}) \\
Pop~III & \begin{tabular}[t]{@{}l@{}}$1.12\times10^{-23}$\\$f_{\rm ref}=0.1\,{\rm Hz}$\end{tabular} & $9.08\times10^{-11}$ & Set $\rho_{\rm III}=2.00\times10^5\,M_\odot\,{\rm Mpc}^{-3}$ (Eq.~\eqref{eq:popiii-ul}) \\
sBBH & \begin{tabular}[t]{@{}l@{}}$5.14\times10^{-25}$\\$f_{\rm ref}=25\,{\rm Hz}$\end{tabular} & $4.65\times10^{-9}$ & \leavevmode Use the sBBH rate in the upper limit column of Table~\ref{tab:lvk-rates}; set $\chi_{1,2}=+0.7$ and $\epsilon_{\rm gw}=0.08$ (Eq.~\eqref{eq:bbh-ceiling}) \\
\bottomrule
\end{tabularx}
\end{table*}

\subsubsection{Pop~III BBHs (decihertz Band)}
\label{sec:POPIII}
The Pop~III background is constrained by the scarcity of metal-free gas in the early Universe. These first generation stars formed at high redshift, and their reservoir is the historical mass formed in metal-free stars~\cite{Inayoshi2020,Kinugawa2014}. These  stars can leave unusually massive black holes, making them easier to detect at lower frequencies than typical sBBHs. Formation models span isolated binary evolution and dynamical hardening, leaving merger fractions, delay times, redshifts, and spectra uncertain~\cite{Kinugawa2014,Belczynski2017,Santoliquido2023, Liu2024PopIII}. Magnetic fields and radiative feedback also make the Pop~III stellar mass distribution uncertain ~\cite{2025MNRAS.541L...1S,2026arXiv260400197C}.

For the benchmark, we assume that $1\%$ of the Pop~III stellar mass merges in nonspinning $50+50\,M_\odot$ binaries. We model each binary with a nonspinning IMR energy spectrum for ${\rm d}E/{\rm d}f_r$ in Eq.~\eqref{eq:cosmo-integral}.  This spectrum sets $\epsilon_{\rm gw}=0.05$ in Eq.~\eqref{eq:global-energy} and $\mathcal H(f)$ in Eq.~\eqref{eq:history-spectrum-factor}. This corresponds to a merger efficiency of $10^{-4}\,M_\odot^{-1}$, within the range found by \citet{Liu2024PopIII}.  We assume that Pop~III BBHs merge at the same redshift at which their progenitor stars form, so the merger rate history follows the Pop~III star-formation history of \citet{Liu2020}. For this channel, $\rho_{\rm res}=\rho_{\rm III}$, $\rho_{\rm src}^{\rm III}$ is the total binary mass entering mergers, and $f_{\rm merge}^{\rm III}=\rho_{\rm src}^{\rm III}/\rho_{\rm III}$ is this mass divided by the historical Pop~III stellar mass.

The source mass is
\begin{equation}
\begin{aligned}
\rho_{\rm src}^{\rm III}
&=f_{\rm merge}^{\rm III}\rho_{\rm III}\\
&=7.33\times10^2\,M_\odot\,{\rm Mpc}^{-3}
\left(\frac{\rho_{\rm III}}{\rho_{\rm III,bench}}\right)
\left(\frac{f_{\rm merge}^{\rm III}}{0.01}\right).
\end{aligned}
\label{eq:popiii-source-density}
\end{equation}
For this benchmark, the mass-weighted mean merger redshift is 8.70, and $\mathcal H_{\rm III}\equiv\mathcal H(0.1\,{\rm Hz})$. Equation~\eqref{eq:universal-scaling} gives
\begin{equation}
\begin{aligned}
&h_c^{\mathrm{bench}}(0.1\,{\rm Hz})=6.80\times10^{-24}\\
&\times\left(\frac{\rho_{\rm III}}{\rho_{\rm III,bench}}\right)^{1/2}
\left(\frac{f_{\rm merge}^{\rm III}}{0.01}\right)^{1/2}
\left(\frac{\epsilon_{\rm gw}}{0.05}\right)^{1/2}
\left(\frac{\mathcal H}{\mathcal H_{\rm III}}\right)^{1/2}.
\end{aligned}
\label{eq:popiii-benchmark}
\end{equation}
The benchmark background peaks at $\Omega_{\rm gw}=1.65\times10^{-11}$ near $15\,{\rm Hz}$. To set an upper limit, we ask how large the background could be if the total Pop~III stellar mass reached the maximum allowed by reionization. In the reionization model of \citet{Inayoshi2021}, constrained by the Planck optical depth~\citep{Planck2020}, the cumulative Pop~III stellar mass density is limited to $\rho_{\rm III,UL}=2.00\times10^5\,M_\odot\,{\rm Mpc}^{-3}$. We use this value while keeping $f_{\rm merge}=0.01$, the merger history, and the spectrum fixed, so only the available stellar mass changes:
\begin{equation}
\begin{aligned}
&h_c^{\mathrm{UL}}(0.1\,{\rm Hz})=1.12\times10^{-23}\\[-0.3ex]
&\times\left(\frac{\rho_{\rm III}}{\rho_{\rm III,UL}}\right)^{1/2}
\left(\frac{f_{\rm merge}^{\rm III}}{0.01}\right)^{1/2}\left(\frac{\epsilon_{\rm gw}}{0.05}\right)^{1/2}
\left(\frac{\mathcal H}{\mathcal H_{\rm III}}\right)^{1/2}.
\end{aligned}
\label{eq:popiii-ul}
\end{equation}

\subsubsection{sBBHs (LVK Band)}
\label{sec:stellarBBH}
sBBH mergers across cosmic history are expected to produce a GWB in the LVK band~\citep{2016PhRvX...6c1018C,LVKO4aStoch}.  Some sBBH remnants may undergo further mergers in AGN disks, grow into IMBHs, and enter the AGN--IMRI channel of Sec.~\ref{sec:IMBH-SMBHcapture}~\cite{Mingarelli2026b}; here we calculate the background from the sBBH mergers themselves.  We use the GWTC-5.0 merger rate in Table~\ref{tab:lvk-rates} to determine the total binary mass processed through this channel, $f_{\rm merge}^{\rm BBH}\rho_\star$, and then apply Eq.~\eqref{eq:universal-scaling} at $25\,{\rm Hz}$.  The central rate defines the benchmark; the upper rate and spectrum with aligned spins define the upper limit.

To convert the merger rate into source mass, we use $\langle M_{\rm BBH}\rangle=21.69\,M_\odot$ as a representative value, obtained from our approximation to the central GWTC-5.0 mass and mass ratio distributions. We hold the source-frame rate constant for $T_{\rm eff}=10$ Gyr.

For this calculation, $\rho_{\rm res}=\rho_\star$, $\rho_{\rm src}^{\rm BBH}$ is the total binary mass entering mergers~\citep[cf.][]{SchiebelbeinZwack2024}, and $f_{\rm merge}^{\rm BBH}=\rho_{\rm src}^{\rm BBH}/\rho_\star$.  The source-mass density is then
\begin{equation}
\begin{aligned}
&\rho_{\rm src}^{\rm BBH}=f_{\rm merge}^{\rm BBH}\rho_\star
=6.59\times10^3\,M_\odot\,{\rm Mpc}^{-3}\\
&\times\!\left(\frac{R_{\rm BBH}}{30.4\,{\rm Gpc}^{-3}{\rm yr}^{-1}}\right)\!\!
\left(\frac{\langle M_{\rm BBH}\rangle}{21.69\,M_\odot}\right)\!\!
\left(\frac{T_{\rm eff}}{10\,{\rm Gyr}}\right)\, .
\end{aligned}
\label{eq:bbh-source-density}
\end{equation}
With $\rho_{\star,\rm bench}=6.30\times10^8\,M_\odot\,{\rm Mpc}^{-3}$, the benchmark source fraction is $f_{\rm merge,bench}^{\rm BBH}=1.05\times10^{-5}$. Across the benchmark rate interval in Table~\ref{tab:lvk-rates}, $\rho_{\rm src}^{\rm BBH}=(6.59^{+1.41}_{-1.17})\times10^3\, M_\odot\,{\rm Mpc}^{-3}$ and $f_{\rm merge}^{\rm BBH}=(1.05^{+0.22}_{-0.19})\times10^{-5}$, with the representative mass and history fixed.

For the benchmark, we use a nonspinning, equal-mass binary and $\epsilon_{\rm gw}=0.05$, as specified in Sec.~\ref{sec:energylimits}.  Let $\mathcal H_{\rm BBH}$ denote $\mathcal H(25\,{\rm Hz})$ for this benchmark.  From Eq.~\eqref{eq:universal-scaling}:
\begin{equation}
\begin{aligned}
&h_c^{\mathrm{bench}}(25\,{\rm Hz})=4.03\times10^{-25}\\
&\times\!\left(\!\!\frac{\rho_\star}{\rho_{\star,\rm bench}}\!\!\right)^{\!1/2}\!\!
\left(\!\!\frac{f_{\rm merge}^{\rm BBH}}{1.05\times10^{-5}}\!\!\right)^{\!1/2}\!\!
\left(\frac{\epsilon_{\rm gw}}{0.05}\right)^{\!1/2}\!\!
\left(\!\frac{\mathcal H}{\mathcal H_{\rm BBH}}\!\right)^{\!1/2}\mathrlap{.}
\end{aligned}
\label{eq:bbh-benchmark}
\end{equation}

For the upper limit, we change only the merger rate and spin. The sBBH rate in the upper limit column of Table~\ref{tab:lvk-rates} gives $f_{\rm merge}^{\rm BBH}=1.70\times10^{-5}$.  Setting $\chi_1=\chi_2=0.7$ raises the radiated fraction to $\epsilon_{\rm gw}=0.08$~\cite{Husa2016}. The rate change produces most of the increase at $25\,{\rm Hz}$:
\begin{equation}
\begin{aligned}
&h_c^{\mathrm{UL}}(25\,{\rm Hz})=5.14\times10^{-25}\\
&\times\!\left(\!\!\frac{\rho_\star}{\rho_{\star,\rm bench}}\!\!\right)^{\!1/2}\!\!
\left(\!\!\frac{f_{\rm merge}^{\rm BBH}}{1.70\times10^{-5}}\!\!\right)^{\!1/2}\!\!
\left(\frac{\epsilon_{\rm gw}}{0.08}\right)^{\!1/2}\!\!
\left(\!\!\frac{\mathcal H}{\mathcal H_{\rm BBH,UL}}\!\!\right)^{\!1/2}\mathrlap{.}
\end{aligned}
\label{eq:bbh-ceiling}
\end{equation}

Appendix~\ref{app:bbh-lvk-comparison} places these strain values in the same $\Omega_{\rm gw}(25\,{\rm Hz})$ convention used by LVK and converts the LVK stochastic upper limit into conditional bounds on processed mass and merger rate.  It also identifies the population moment constrained by the stochastic search and explains why our calculations need not reproduce a full-population inference. Although the sBBH and AGN--IMRI backgrounds arise from distinct GW events and their energy densities therefore add, both draw on the same black hole mass reservoir, so mass assigned to one channel cannot also be counted in the other.

\subsection{Double white dwarfs in the LISA band}
\label{sec:DWD}

DWDs produce two distinct signals in the LISA band: a local, anisotropic Galactic foreground~\cite{Robson2019,BreivikMingarelliLarson2020} and an approximately isotropic cosmological background from DWDs in other galaxies~\citep{Farmer2003,Boileau2025}.  As LISA trails the Earth in a heliocentric orbit, its response to the Galactic foreground varies annually, helping distinguish it from approximately isotropic backgrounds~\citep{Mingarelli2026b}.

\citet{Farmer2003} found $\Omega_{\rm gw}(1\,{\rm mHz})=(1$--$6)\times10^{-12}$ for the extragalactic DWD background.  More recent population synthesis calculations show that its amplitude depends strongly on cosmic star formation and binary evolution~\citep{HofmanNelemans2024}. Using COSMIC~\citep{2020ApJ...898...71B}, \citet{Boileau2025} find that it should be detectable by LISA.

We do not include the extragalactic DWD background among the six benchmarks modeled here --- Figure~\ref{fig:gwb-benchmarks} shows the Galatic foreground for comparison only.  A benchmark in our framework would require determining how much stellar mass forms DWDs, their mass distribution and cosmic history, and how they evolve through contact and mass transfer.  At the published amplitudes, adding this component would not change the SMBHB-dominated integrated sum at the precision quoted in Sec.~\ref{sec:astroCeilingOmegaGW}.  We therefore leave a benchmark and upper limit in this framework to future work.

\section{Integrated GWB Energies}
\label{sec:astroCeilingOmegaGW}

Integrating over frequency places backgrounds in different frequency bands on the same energy scale.  Each $\mathcal E_i$ is already integrated over frequency, so we can add the six benchmark energies in Table~\ref{tab:population_params} without constructing a composite spectrum:
\begin{equation}
\mathcal E_{\rm bench}^{(6)}
\equiv\sum_i\mathcal E_{i,{\rm bench}}
=\left(2.04^{+0.90}_{-0.56}\right)\times10^{-7}.
\label{eq:benchmark-energy-sum}
\end{equation}
This sum compares six separately constructed benchmarks, but it is not a prediction from a single population model.  Such a prediction could be constructed by generating all six channels from a common cosmic history, tracking mass that participates in successive GW events, and adding the resulting GW energies.  We leave such a model for future work.  This sum excludes both DWD signals discussed in Sec.~\ref{sec:DWD}, which are not among the six benchmarks.  For SMBHBs, we use the integrated energy from Eq.~\eqref{eq:circular-inspiral-energy}.  Although the sBBH and AGN--IMRI channels share mass, the sBBH mergers that assemble an IMBH and its later capture by an SMBH are distinct GW events, so their energies add.

Using the benchmark values in Table~\ref{tab:population_params}, SMBHBs contribute 98.9\% of the sum in Eq.~\eqref{eq:benchmark-energy-sum}, sBBHs 0.9\%, and the other four populations together 0.2\%. The quoted range includes the LM24 SMBH-density range and the GWTC-5.0 rate intervals, with the BBH rate varied jointly in the AGN--IMRI and sBBH terms. The GWTC-5.0 merger-rate intervals in Table~\ref{tab:lvk-rates} change the sum by only \(+0.35\%/-0.23\%\), much less than the change produced by the LM24 range for $\rho_\bullet$.

For a second energy sum, we use the integrated energies $\mathcal E_i$ listed in Table~\ref{tab:upper_limits}.  For the SMBHB term, we use the one-generation merger ($q=1,\ z\to0$) value in Table~\ref{tab:upper_limits}. The factor \(2.70\) applies only to \(h_c^2\) at a fixed frequency and is not included in the integrated energy (Appendix~\ref{app:szq24-check}). The reference sum is
\begin{equation}
\mathcal E_{\rm ref}^{(6)}
\equiv\sum_i\mathcal E_i
=\left(3.33^{+1.32}_{-0.83}\right)\times10^{-7}.
\label{eq:integrated-reference-sum}
\end{equation}
SMBHBs again dominate the sum, contributing to $89\%$ of the total. We emphasize that this a reference sum, and not a global ceiling, since the six cases need not occur together. However, such a ceiling is indeed interesting and is the subject of future work.

\section{Discussion}
\label{sec:discussion}

We derive a strain ceiling of $h_c^{\rm ceil}(1\,{\rm yr}^{-1})= 3.76^{+0.76}_{-0.56}\times10^{-15}$ (Eq.~\eqref{eq:smbhb-repeated-ceiling}), based on the LM24 SMBH census and merger hierarchy from SZQ24. This ceiling encompasses the published nanoHertz GWB amplitudes. We can invert the ceiling framework with a GWB measurement to infer the underlying properties of the source population. For example, using the revised NANOGrav GWB measurement~\citep{Larsen2026,Agarwal2026}, combined with the LM24 census, we can infer $f_{\rm merge}=1.11^{+1.35}_{-0.63}$ (90\%; Eq.~\eqref{eq:nanograv-fmerge}), consistent with one merger per present-day remnant. This estimate assigns the full measured amplitude to SMBHBs, however, pulsar noise may still play some role ~\citep{Hazboun2020,Goncharov2021,Larsen2024,Hazboun2025,Larsen2026}.

The calculation builds on the energy relation derived by \citet{Phinney2001}.  \citet{ZhuCuiThrane2019} used the local SMBH mass function to calculate a maximum SMBHB GWB amplitude, assuming that every present-day SMBH was produced by a $q=1$ merger. SZQ24 added earlier merger generations and showed that their contribution to the strain converges; Appendix~\ref{app:szq24-check} compares their calculation with ours.  LM24 showed that their inferred local SMBH population predicts a PTA-compatible amplitude ~\citep{Liepold2024,LiepoldMa2026Erratum}.

PTA measurements can now constrain models of high-mass SMBH growth. The revised NANOGrav GWB amplitude is approximately equal to the idealized SZQ24 VDF-census limit of \(2.05\times10^{-15}\), but lies below the LM24 ceiling of \(3.76\times10^{-15}\) (Table~\ref{tab:szq-ladder}). With the VDF census, reproducing the revised NANOGrav GWB amplitude requires a merger hierarchy close to the infinite limit, whereas LM24 implies fewer mergers per remnant. The quasar luminosity history provides a second constraint. Matching the LM24 local mass density to the observed quasar luminosity history requires lower radiative efficiencies than standard luminous accretion models predict~\citep{2025PhRvD.112l3018S}. Combined with local SMBH censuses~\citep{2025arXiv250908041S} and the quasar luminosity history, the nanoHertz GWB amplitude constrains merger, accretion, and AGN feedback models for high-mass SMBH growth~\citep{Tillman2026}. Further independent censuses and better-controlled SMBH mass measurements can determine whether the GWB reflects a larger high-mass population, more mergers per remnant, or different accretion and feedback histories.

Since $h_c^2(f)\propto\mathcal{M}_c^{5/3}$, the abundance of massive binaries affects not only the GWB amplitude but also its anisotropy, departures from a smooth strain power law, and the likelihood that individual binaries are detectable~\citep{JaffeBacker2003,2025ApJ...978...31A,Mingarelli2013,Gair2014}. The NANOGrav free spectrum already disfavors the most extreme few-source models~\citep{2025PhRvD.111b3043S}, and  continuous GW searches test the high-mass population directly via all-sky and targeted searches~\citep{Agazie2023CW,Antoniadis2024CW, Zhao2025CW,Agarwal2025,2026PhRvD.114d4055M, CaseyClyde2025}.

The other five backgrounds show how the same calculation uses different astrophysical constraints: LVK rates for BNSs and sBBHs; the LVK BBH rate, retention, and delivery fractions for AGN--IMRIs; the compact object supply in NSCs for EMRIs; and the historical metal-free stellar mass for Pop~III.

The EMRI and sBBH cases allow direct comparisons with existing population calculations. For EMRIs, translating the \citet{Babak2017} capture rate into the mass captured from NSCs gives a benchmark within about $30\%$ of the detailed population calculation of \citet{BonettiSesana2020} for the same rate model.  Appendix ~\ref{app:bbh-lvk-comparison} recovers the standard LVK sBBH population integral and expresses the stochastic limit as a bound on the mass processed through mergers.  The inverse calculation also provides a cross-band test: LVK constrains the mass entering sBBH mergers, while a LISA measurement or upper limit could constrain the fraction of that mass retained in IMBHs and later delivered to central SMBHs~\citep{Mingarelli2026b}.

Integrating over frequency places the six backgrounds on the same energy scale.  SMBHBs contribute $98.9\%$ of $\mathcal E_{\rm bench}^{(6)}$ in Eq.~\eqref{eq:benchmark-energy-sum} and $89\%$ of the reference sum in Eq.~\eqref{eq:integrated-reference-sum}. This dominance reflects the mass processed through mergers: the measured BNS and sBBH rates place only a small fraction of the stellar reservoir in merging binaries, whereas the SMBHB benchmark assigns one merger generation to the local SMBH census.  Neither sum is a global ceiling, because the six cases are constructed separately and connected channels can share mass.

The SMBHB result is a physical ceiling because the local census independently bounds the source mass.  The other five results remain upper limits because no comparable census constrains both the abundance and mass distribution of their sources.  A global astrophysical ceiling would require a joint population model that follows shared mass through successive GW events.  With an independent source census, the same calculation could set a ceiling for additional channels, including the extragalactic DWD background~\citep{Boileau2025,2026arXiv260505308K}. Early-Universe backgrounds from phase transitions, cosmic strings, or primordial black holes are not limited by stellar and black-hole reservoirs and require separate cosmological bounds~\citep{Caprini2018,Auclair2020,Carr2021, GiblinThrane2014,2024arXiv240915572W}. Together, GWB measurements and independent source censuses turn stochastic signals into quantitative probes of the mass processed through mergers across cosmic history.

\section{Conclusion}
\label{sec:conclusion}
The LM24 census contains enough high-mass SMBHs to produce the nanoHertz GWB. No additional SMBH population is required. Indeed, the census bounds the GWB, while the measured GWB determines the merger factor. Applying this mass-based accounting to six populations and integrating over frequency places their backgrounds on a common energy scale. SMBHBs dominate both energy sums because far more mass is processed through their mergers.  The other five results remain theoretical upper limits because comparable source censuses do not yet exist; improved censuses could make physical ceilings possible in other bands.

If no viable astrophysical source budget can support a measured background, the signal would point to an unmodeled source population or new physics.
\appendix

\section{Relation to Sato-Polito et al. (2024)}
\label{app:szq24-check}

SZQ24 used the framework of \citet{Phinney2001} to connect the PTA background to a local SMBH census.  Evaluated with their inputs, Eq.~\eqref{eq:cosmo-integral} reduces to their Eq.~(15).  The two treatments make similar assumptions: that the present-day SMBH mass is identified with the total mass of its progenitor binary, neglecting GW losses and subsequent accretion.  The assignment of one merger to each present-day SMBH corresponds to our $f_{\rm merge}=1$.

The principal distinction is the adopted SMBH census and population model. SZQ24 construct SMBH mass functions by combining galaxy stellar VDFs with the $M_\bullet$--$\sigma$ relations of \citet{MM13} and \citet{Kormendy2013}.  Their fiducial model pairs the VDF of \citet{Bernardi2010} with the $M_\bullet$--$\sigma$ relation fitted by \citet{MM13} to their full galaxy sample.  We instead adopt the stellar mass based local census of \citet{Liepold2024,LiepoldMa2026Erratum}.

For a direct numerical comparison, we evaluated Eq.~\eqref{eq:cosmo-integral} at the central SZQ24 parameter values, analytically averaging over the reported $0.38$-dex intrinsic scatter and adopting their mass ratio and redshift distributions.  We obtain $\rho_\bullet=4.54\times10^5\,M_\odot\,{\rm Mpc}^{-3}$, in agreement with their quoted $(4.6\pm0.9)\times10^5\,M_\odot\,{\rm Mpc}^{-3}$.  Our corresponding amplitude, $A_{\rm yr}=9.95\times10^{-16}$, agrees at the ${\sim}10\%$ level with the $1.11\times10^{-15}$ value implied by their Eq.~(21). Table~\ref{tab:szq-ladder} summarizes the resulting amplitudes for the two populations.  The two LM24 one-generation entries use the same density given in Eq.~\eqref{eq:rho_bullet_result}.

\begin{table}[b!]
\caption{\label{tab:szq-ladder}Parallel SMBHB constructions using the SZQ24 and LM24 populations.  The one-generation maximum sets $q=1$ and $z\to0$; the final row adds the mass based hierarchy developed by SZQ24.  Entries are $A_{\rm yr}\times 10^{-15}$.}
\begin{ruledtabular}
\begin{tabular}{lcc}
Construction & SZQ24 &  This Work \\
\colrule
Fiducial one generation & 0.99 & 2.11 \\
One-generation maximum & 1.25 & 2.29 \\
Mass-conserving hierarchy & 2.05 & 3.76 \\
\end{tabular}
\end{ruledtabular}
\end{table}

For the SZQ24 population, the equal-mass and zero-redshift specializations raise $h_c^2$ by factors of 1.4 and 1.1, respectively.  As SZQ24 show, the mass-conserving hierarchy of earlier generations supplies a further factor of 2.70 at fixed frequency.  Each earlier generation contributes less because its binaries are less massive.  We apply the hierarchy factor to the physical ceiling in Eq.~\eqref{eq:smbhb-repeated-ceiling}, but not to the one-generation energy integral in Eq.~\eqref{eq:integrated-reference-sum}.

The amplitudes in the two columns differ because of the SMBH census and the adopted mass ratio and redshift distributions; the underlying energy relation is the same.  Within their velocity-dispersion census, SZQ24 found that realistic merger histories could not fully account for the PTA amplitude. Later work showed that the higher stellar mass census used here reduces the need for repeated mergers~\citep{2025PhRvD.112l3018S}.

SZQ24 ask in their title, ``Where are the supermassive black holes measured by PTAs?''  In the LM24 census, they are already present at the observed high-mass end of the local SMBH population; no additional population is required.

\section{Relation to LVK GWB calculations}
\label{app:bbh-lvk-comparison}

LVK calculates the sBBH background by integrating the GW energy emitted by mergers over their masses, mass ratios, spins, and redshifts.  Substituting $N(z)=R(z)/[(1+z)H(z)]$ and $f_r=(1+z)f$ into Eqs.~\eqref{eq:omega-strain} and \eqref{eq:cosmo-integral} gives the standard rate-density form used by LVK~\citep{Phinney2001,Regimbau2011,LVKO4aStoch},
\begin{equation}
\Omega_{\rm gw}(f)
=\frac{f}{\rho_c}
\int_0^{z_{\rm max}}
\frac{R(z)}{(1+z)H(z)}
\left\langle\frac{{\rm d}E}{{\rm d}f_r}\right\rangle_z
{\rm d}z,
\label{eq:lvk-master}
\end{equation}
where the brackets denote an average over the binary population at each redshift, and the energy spectrum is evaluated at $f_r=(1+z)f$.

For our mass accounting, we also weight the same events by their total binary mass:
\begin{equation}
\rho_{\rm src}^{\rm BBH}
=\int_0^{z_{\rm max}} R(z)
\left|\frac{{\rm d}t}{{\rm d}z}\right|
\left\langle M_{\rm tot}\right\rangle_z\,{\rm d}z .
\label{eq:lvk-bbh-source-mass}
\end{equation}
This is the mass-weighted counterpart of Phinney's remnant number density $N_0=\int N(z)\,{\rm d}z$.  We write $\rho_{\rm src}^{\rm BBH}=f_{\rm merge}^{\rm BBH}\rho_\star$ and define $\epsilon_{\rm gw}$ as the processed-mass-weighted mean fraction radiated in GWs.  Equations~\eqref{eq:omega-strain} and \eqref{eq:history-spectrum-factor} then give $\rho_c\Omega_{\rm gw}(f)= \epsilon_{\rm gw}\rho_{\rm src}^{\rm BBH}\mathcal H(f)$ by definition. Substituting $\rho_{\rm src}^{\rm BBH}=f_{\rm merge}^{\rm BBH}\rho_\star$ into this identity and combining it with Eq.~\eqref{eq:omega-strain} gives Eq.~\eqref{eq:universal-scaling}.  This rewriting introduces no approximation.

Our benchmark makes three simplifying choices: a constant merger rate for 10~Gyr, a representative total mass, and a nonspinning, equal-mass inspiral--merger--ringdown energy spectrum.  LVK instead averages over the inferred distributions of masses, mass ratios, spins, and redshifts.  In the leading-order inspiral regime, a circular, GW-driven binary has the energy spectrum ~\citep{1963PhRv..131..435P,1964PhRv..136.1224P,Phinney2001}
\begin{equation}
\frac{{\rm d}E}{{\rm d}f_r}
= \frac{\pi^{2/3}}{3}\,\mathcal{M}_c^{5/3}\,f_r^{-1/3}.
\label{eq:single-spectrum}
\end{equation}
Inserting Eq.~\eqref{eq:single-spectrum} into Eq.~\eqref{eq:lvk-master} gives
\begin{align}
\Omega_{\rm gw}(f)&\propto f^{2/3}
\int_0^{z_{\rm max}} {\rm d}z
\frac{R(z)\left\langle\mathcal{M}_c^{5/3}\right\rangle_z}
{(1+z)^{4/3}H(z)} .
\label{eq:lvk-bbh-mass-moment}
\end{align}
This is the rate-density form of the familiar $f^{2/3}$ inspiral result and generalizes Eq.~(10) of \citet{Phinney2001} to a mass distribution. Equation~\eqref{eq:lvk-bbh-mass-moment} weights the population by $\mathcal{M}_c^{5/3}$, so the relevant mass moment depends on the full mass and mass-ratio distributions rather than on a single mean total mass.  It also weights the merger rate by $(1+z)^{-4/3}H(z)^{-1}$.  Our representative mass, constant-rate benchmark therefore need not reproduce the LVK full-population result.

The LVK O1--O4a stochastic search for a fixed $\Omega_{\rm gw}\propto f^{2/3}$ spectrum gives $\Omega_{\rm gw}(25\,{\rm Hz})<2.0\times10^{-9}$ at 95\% credibility with a log-uniform amplitude prior~\cite{LVKO4aStoch}.  Assigning this entire allowed component to sBBHs and applying our framework with the representative binary and 10 Gyr constant rate history of Sec.~\ref{sec:stellarBBH} gives
\begin{equation}
\begin{aligned}
f_{\rm merge}^{\rm BBH}&<1.5\times10^{-4},\\
\rho_{\rm src}^{\rm BBH}&<9.4\times10^4\,M_\odot\,{\rm Mpc}^{-3},\\
R_{\rm BBH}&<4.3\times10^2\,{\rm Gpc}^{-3}{\rm yr}^{-1}.
\end{aligned}
\label{eq:lvk-bbh-observational-bounds}
\end{equation}

The three quantities in Eq.~\eqref{eq:lvk-bbh-observational-bounds} are equivalent conditional forms of a single 95\%-credible limit and not independent constraints. In fact, a catalog inference of the analogous cumulative component-mass density has already been presented by \citet{SchiebelbeinZwack2024}.

The comparisons answer distinct questions.  A catalog prediction asks what background the resolved population should produce; our theoretical upper limit asks how large the sBBH background can be within the adopted input ranges; and the stochastic limit asks how large a fixed $f^{2/3}$ component the cross-correlation data allow.  Our inverse mapping asks what processed mass and merger rate would produce that allowed component, without adding information to the LVK likelihood.  For our representative model, the mapped stochastic rate bound would equal the $R_{\rm BBH}=49.4\,{\rm Gpc}^{-3}{\rm yr}^{-1}$ upper-end rate adopted for our theoretical upper limit in Table~\ref{tab:lvk-rates} when $\Omega_{\rm gw}(25\,{\rm Hz})\simeq2.3\times10^{-10}$, about a factor of nine below the O1--O4a limit.  The stochastic channel constrains the mass- and redshift-weighted moment in Eq.~\eqref{eq:lvk-bbh-mass-moment}---as an upper bound now and a measurement after detection---including unresolved mergers. The catalog and stochastic data are therefore complementary and belong in a joint population likelihood~\citep{Callister2020,LVKO4aStoch}.

{\section*{Software}} The following software was used for this work: NumPy~\cite{Harris2020NumPy}, SciPy~\cite{Virtanen2020SciPy}, Matplotlib~\cite{Hunter2007Matplotlib}, OpenAI's ChatGPT~\cite{OpenAIChatGPT} and Anthropic's Claude Code~\cite{AnthropicClaudeCode}. The codes developed for this work are publicly available at \url{https://github.com/ChiaraMingarelli/gwb_ceilings}. An interactive plotting tool for exploring the ceilings and benchmarks is available at \url{https://gwbceilings.streamlit.app}.

\vspace{2mm}

\begin{acknowledgments}
C.\,M.\,F.\,M. is grateful to B. Burkhart, A. Sesana, S. E. K. Ford, B. McKernan, P. Duffell, Q. Zheng, B. Larsen, A. Brazier, and V. Ozolins for useful comments. She also thanks P. Natarajan for helpful discussions and for noting related, independent interest in mass-dependent estimates of $\Omega_{\rm BH}$ within her group. This work was supported in part by the National Science Foundation under Grant PHY--2020265, and NASA's LISA Preparatory Science program under Grant 80NSSC24K0440.
\end{acknowledgments}

\bibliographystyle{apsrev4-2}
\bibliography{bib}

\end{document}